\documentclass[a4paper,11pt]{article}
\usepackage{pos}
\usepackage{amsmath}
\usepackage{braket}
\usepackage{array,multirow}
\usepackage{tikz}
\usepackage{dsfont}
\usetikzlibrary{snakes}
\usetikzlibrary{math}
\usetikzlibrary{calc}
\tikzset{decoration={snake,amplitude=.2mm,segment length=1mm,
		post length=0mm,pre length=0mm}}

\newcommand{\J}{{\tilde{J}}}
\newcommand{\MatrixTwoXTwo}[4]{	\left( \begin{array}{cc} #1 & #2 \\ 
		#3 & #4 \\ \end{array}\right)}
\newcommand{\MatrixThreeXThree}[9]{	\left( \begin{array}{ccc} #1 & #2 & #3 \\ 
		#4 & #5 & #6 \\ 
		#7 & #8 & #9 \\\end{array}\right)}
\newcommand{\xx}[2]{\chi_{#1\rightarrow #2}(r)}
\newcommand{\Vector}[3]{\left( \begin{array}{c} #1 \\ #2 \\ #3 \\ \end{array}\right)}

\title{Bottomonium resonances from lattice QCD static-static-light-light potentials}

\author*[a]{Lasse Mueller}
\author[b]{Pedro Bicudo}
\author[b]{Nuno Cardoso}
\author[a,c]{Marc Wagner}

\affiliation[a]{Johann Wolfgang Goethe-Universit\"at Frankfurt am Main, Institut f\"ur Theoretische Physik, \\ Max-von-Laue-Stra{\ss}e 1, D-60438 Frankfurt am Main, Germany}

\affiliation[b]{CeFEMA, Dep.\ F\'{\i}sica, Instituto Superior T\'ecnico, Universidade de Lisboa, \\ Av.\ Rovisco Pais, 1049-001 Lisboa, Portugal}

\affiliation[c]{Helmholtz Research Academy Hesse for FAIR, Campus Riedberg, Max-von-Laue-Stra{\ss}e 12, D-60438 Frankfurt am Main, Germany}

\emailAdd{lmueller@itp.uni-frankfurt.de}
\emailAdd{bicudo@tecnico.ulisboa.pt}
\emailAdd{nuno.cardoso@tecnico.ulisboa.pt}
\emailAdd{mwagner@itp.uni-frankfurt.de}

\abstract{We study $I=0$ quarkonium resonances decaying into pairs of heavy-light mesons using static-static-light-light potentials from lattice QCD. To this end, we solve a coupled channel Schr\"odinger equation with a confined quarkonium channel and channels with a heavy-light meson pair to compute phase shifts and $\mbox{T}$ matrix poles for the lightest decay channel. We discuss our results for $S$, $P$, $D$ and $F$ wave states in the context of corresponding experimental results, in particular for $\Upsilon(10753)$ and $\Upsilon(10860)$.}

\FullConference{%
 The 38th International Symposium on Lattice Field Theory, LATTICE2021
  26th-30th July, 2021
  Zoom/Gather@Massachusetts Institute of Technology
}

\begin{document}
\maketitle

\section{Introduction}

	In this work we study $I=0$ quarkonium resonances using lattice QCD string breaking potentials computed in Ref.\ \cite{Bali:2005fu}. Our approach is based on the diabatic extension of the Born-Oppenheimer approximation and the unitary emergent wave method. In the first step of the Born-Oppenheimer approximation the two heavy quarks are considered as static to compute their potentials, possibly in the presence of two light quarks. In the second step these potentials are used in a coupled channel Schr\"odinger equation describing the dynamics of the heavy quarks \cite{Born:1927}.

	In the past this approach was successfully applied to compute resonances for $\bar b \bar b q q$ systems, where $q$ denotes a light quark of flavor $u$ or $d$ \cite{Bicudo:2017szl}. In this work we investigate $\bar b b$ and $\bar b b \bar q q$ systems, which are technically more complicated, because there are a confined and two meson-meson decay channels. Studying this system is of interest also from an experimental point of view, because corresponding experimental results are available, e.g.\ for $\Upsilon(nS)$, $\Upsilon(10860)$ and $\Upsilon(11020)$.

	We note that there are also ongoing efforts to study the related $I = 1$ system \cite{Prelovsek:2019ywc,Peters:2017hon}.


\section{Coupled channel Schr\"odinger equation}

	We study $\bar Q Q$ and $\bar Q Q \bar q q$ quarkonium systems with $I=0$ and consider the heavy quark spins as conserved quantities. Thus, resulting masses and decay widths will be independent of the heavy quark spins. Such systems can be characterized by the following quantum numbers:
\begin{itemize}
	\item $J^{PC}$: total angular momentum, parity and charge conjugation.
	\item $\J^{PC}$: total angular momentum excluding the heavy $\bar{Q} Q$ spins and corresponding parity and charge conjugation.
	\item $L^{PC}$: orbital angular momentum and corresponding parity and charge conjugation. (For $\bar Q Q$ systems $\J^{PC}$ coincides with $L^{PC}$.)
\end{itemize}

\begin{figure}[htb]
		\centering
		\includegraphics[width=0.7\textwidth]{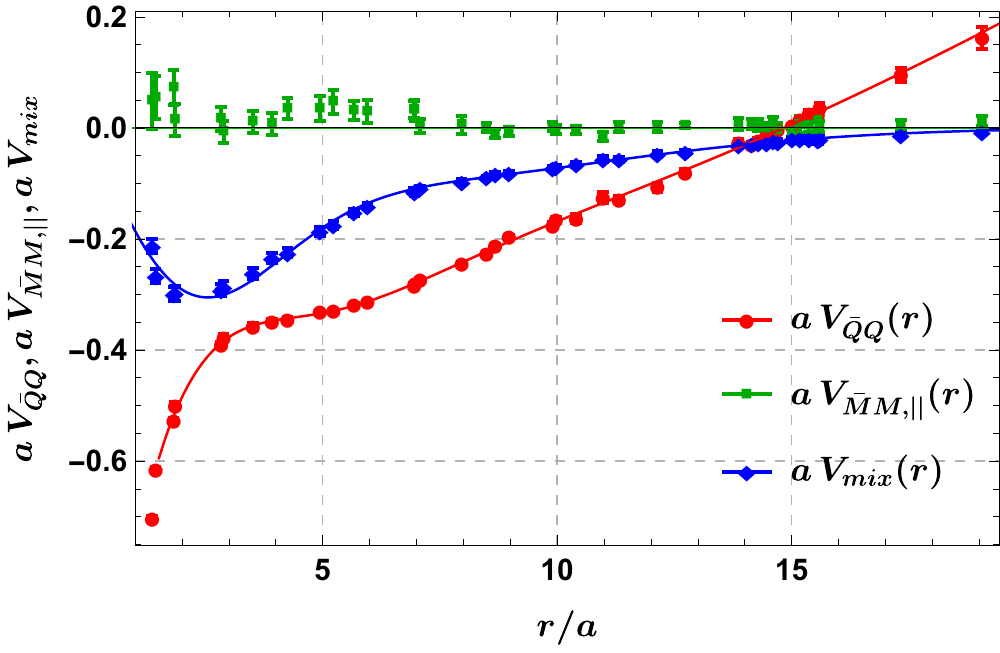}
		\caption{Potentials $V_{\bar{Q}Q}$, $V_{\bar{M}M, \parallel}(r)$ and $V_{\textrm{mix}}(r)$ as functions of $\bar Q Q$ separation $r$. The curves correspond to the parameterizations \eqref{eqn:parameterization1} to \eqref{eqn:parameterization3} with parameters listed in Table \ref{tab:fitsGevFm}.}
		\label{fig:potentials}
	\end{figure}

	In Ref.\ \cite{Bicudo:2019ymo} we derived in detail the Schr\"odinger equation describing these quarkonium systems. In a simplified version, where $s$ quarks are ignored, it is composed of a quarkonium channel $\bar{Q} Q$ and of heavy-light meson-meson channels $\bar{M} M$ with $M = \bar{Q} q$ and $q \in \{ u,d \}$. The wave function has 4 components, $\psi(\mathbf{r}) = (\psi_{\bar{Q} Q}(\mathbf{r}), \vec{\psi}_{\bar{M} M}(\mathbf{r}))$. The first component $\psi_{\bar{Q} Q}(\mathbf{r})$ represents the $\bar Q Q$-channel, while the three lower components $\vec{\psi}_{\bar{M} M}(\mathbf{r})$ represent the spin-1 triplet of the $\bar M M$-channel. In detail the Schr\"odinger equation is given by
	\begin{align}
		\left(-\frac{1}{2}\mu^{-1}\left(\partial_r^2+\frac{2}{r}\partial_r-\frac{\mathbf{L}^2}{r^2}\right)+V(\mathbf{r}) + \MatrixTwoXTwo{E_{\textrm{threshold}}}{0}{0}{2 m_M} -E\right)\psi(\mathbf{r}) = 0,
		\label{eqn:Schroedinger_equation}
	\end{align}
	where $E_{\textrm{threshold}}$ is a constant shift of the confining potential discussed below, $m_M$ denotes the mass of a heavy-light meson, $\mu^{-1} = \textrm{diag}(1/\mu_Q,1/\mu_M,1/\mu_M,1/\mu_M)$ contains the reduced masses of a heavy quark pair and a heavy-light meson pair
	and
	\begin{align}
		V(\mathbf{r}) = \MatrixTwoXTwo{V_{\bar{Q}Q}(r)}{V_{\textrm{mix}}(r)\left(1\otimes\mathbf{e}_r\right)}
		{V_{\textrm{mix}}(r)\left(\mathbf{e}_r\otimes 1\right)}
		{V_{\bar{M}M, \parallel}(r)\left(\mathbf{e}_r\otimes\mathbf{e}_r\right) + 
			V_{\bar{M}M, \perp}(r)\left(1-\mathbf{e}_r\otimes\mathbf{e}_r\right)}.
	\end{align}
	The entries of this potential matrix, $V_{\bar{Q}Q}$, $V_{\textrm{mix}}$, $V_{\bar{M}M, \parallel}$ and $V_{\bar{M}M, \perp}$, are related to static potentials from QCD, which can be computed with lattice QCD, as e.g.\ done in Ref.\ \cite{Bali:2005fu}, where string breaking is studied. Suitable parameterizations are
	\begin{align}
		& V_{\bar{Q}Q}(r) = E_0 - \frac{\alpha}{r} + \sigma r + \sum_{j=1}^{2} c_{\bar{Q}Q, j} \, r \exp\left(-\frac{r^2}{2\lambda^2_{\bar{Q}Q, j}}\right) \label{eqn:parameterization1} \\
		& V_{\textrm{mix}}(r) = \sum_{j=1}^{2} c_{\textrm{mix}, j} \, r \exp\left(-\frac{r^2}{2\lambda^2_{\textrm{mix}, j}}\right) \\
		& V_{\bar{M}M, \parallel}(r) = V_{\bar{M}M, \perp}(r) = 0
		\label{eqn:parameterization3}
	\end{align}
	with parameters listed in Table~\ref{tab:fitsGevFm}. These parameterizations are shown in Figure~\ref{fig:potentials} together with data points, which are based on the lattice QCD results from Ref.\ \cite{Bali:2005fu}. The lattice QCD setup also fixes $E_{\textrm{threshold}}$, which is identical to two times the heavy-light meson mass in that setup.

	\begin{table}[htb]
		\centering
			\begin{tabular}{c|c|c}
				\hline
				potential & parameter & value \\ 
				\hline
				\hline
				$V_{\bar{Q} Q}(r)$ & $E_0$                      & $-1.599(269) \, \textrm{GeV}\phantom{1.^{-1}}$ \\
				& $\alpha$                   & $+0.320(94) \phantom{1.0 \, \textrm{GeV}^{-1}}$ \\
				& $\sigma$                   & $+0.253(035) \, \textrm{GeV}^{2\phantom{-}}\phantom{1.}$ \\
				& $c_{\bar{Q} Q,1}$          & $+0.826(882) \, \textrm{GeV}^{2\phantom{-}}\phantom{1.}$ \\
				& $\lambda_{\bar{Q} Q,1}$    & $+0.964(47) \, \textrm{GeV}^{-1}\phantom{1.0}$ \\
				& $c_{\bar{Q} Q,2}$          & $+0.174(1.004) \, \textrm{GeV}^{2\phantom{-}}$ \\
				& $\lambda_{\bar{Q} Q,2}$    & $+2.663(425) \, \textrm{GeV}^{-1}\phantom{1.}$ \\
				\hline
				\hline
				$V_{\textrm{mix}}(r)$ & $c_{\textrm{mix},1}$       & $-0.988(32) \, \textrm{GeV}^{2\phantom{-}}\phantom{1.0}$ \\
				& $\lambda_{\textrm{mix},1}$ & $+0.982(18) \, \textrm{GeV}^{-1}\phantom{1.0}$ \\
				& $c_{\textrm{mix},2}$       & $-0.142(7) \, \textrm{GeV}^{2\phantom{-}}\phantom{1.00}$ \\
				& $\lambda_{\textrm{mix},2}$ & $+2.666(46) \, \textrm{GeV}^{-1}\phantom{1.0}$ \\
				\hline
			\end{tabular}
		\caption{\label{tab:fitsGevFm}Parameters of the potential parametrizations (\ref{eqn:parameterization1}) to (\ref{eqn:parameterization3}).}
	\end{table}

\section{Scattering matrix for definite $\J$}

	One can expand the wave function $\psi(\mathbf{r})$ in terms of eigenfunctions of $\J$ and project the Schr\"odinger equation to definite $\J$. For $\J>0$ this leads to three coupled ordinary differential equations,
	\begin{eqnarray}
		\nonumber & & \hspace{-0.7cm} \left(-\frac{1}{2}\mu^{-1} \left(\partial^2_r - \frac{1}{r^2} L^2_\J\right) +V_\J(r) + \MatrixThreeXThree{E_{\textrm{threshold}}}{0}{0}{0}{2m_M}{0}{0}{0}{2m_M} - E\right)\Vector{u_{\J}(r)}{\xx{\J-1}{\J}}{\xx{\J+1}{\J}} = \\
		& & = \left( \begin{array}{c} V_{\textrm{mix}}(r) \\ 0 \\ 0 \end{array}\right) \left(\alpha_1 {\J \over 2 \J + 1} r j_{\J-1} (k r) + \alpha_2 {\J+1 \over 2 \J + 1} r j_{\J+1}(k r)\right)
		\label{eqn:coupledChannelSE_3x3}
	\end{eqnarray}
	with $\mu^{-1} = \textrm{diag}(1/\mu_Q,1/\mu_M,1/\mu_M)$, $L_\J^2 = \textrm{diag}(\J(\J+1),(\J-1)\J,(\J+1)(\J+2))$ and
	\begin{eqnarray}
		V_\J(r) &= 
		\left(\begin{array}{ccc}
			V_{\bar{Q} Q}(r) & \sqrt{ \J \over 2 \J+1} V_\textrm{mix}(r)  & \sqrt{\J+1 \over 2\J+1} V_\textrm{mix}(r)  \\
			\sqrt{ \J \over 2 \J+1}  V_\textrm{mix}(r)  & 0 & 0 \\
			\sqrt{ \J+1 \over 2 \J+1}  V_\textrm{mix}(r) & 0  & 0
		\end{array}\right) .
	\end{eqnarray}
	The incident wave can be any superposition of $\bar M M$ with orbital angular momentum $L = \J - 1$ and $L = \J + 1$. For example, an incident pure $\bar M M$ wave with $L = \J - 1$ corresponds to $(\alpha_1,\alpha_2) = (1,0)$. The boundary conditions are as follows:
	\begin{align}
		\label{eqn:boundary_conditions_1}
		u_{\J}(r) &\propto r^{\J+1} , \quad
		\xx{L}{\J} \propto r^{L+1} &&\textrm{for} \quad r \rightarrow 0 \\
		u_{\J}(r) &= 0 &&\textrm{for} \quad r \rightarrow \infty,
	\end{align} 
	for $(\alpha_1,\alpha_2) = (1,0)$
	\begin{align}
		\label{eqn:boundary_conditions_2}
		\xx{\J-1}{\J} = it_{\J-1,\J-1} r h_{\J-1}^{(1)}(kr), \quad
		\xx{\J+1}{\J} = it_{\J-1,\J+1} r h_{\J+1}^{(1)}(kr) \quad \textrm{for} \quad r \rightarrow \infty
	\end{align} 
	and for $(\alpha_1,\alpha_2) = (0,1)$
	\begin{align}
		\label{eqn:boundary_conditions_3}
		\xx{\J-1}{\J} = it_{\J+1,\J-1}\,r\,h_{\J-1}^{(1)}(kr), \quad
		\xx{\J+1}{\J} = it_{\J+1,\J+1}\,r\,h_{\J+1}^{(1)}(kr) \quad \textrm{for} \quad r \rightarrow \infty .
	\end{align} 
	Eqs.\ \eqref{eqn:boundary_conditions_2} and \eqref{eqn:boundary_conditions_3} define $\mbox{S}$ and $\mbox{T}$ matrices,
	\begin{align}
	  \mbox{T}_{\J} = \MatrixTwoXTwo{t_{\J-1,\J-1}}{t_{\J+1,\J-1}}{t_{\J-1,\J+1}}{t_{\J+1,\J+1}}, \quad \mbox{S}_{\J} = 1 + 2 i \mbox{T}_{\J} .
		\label{eqn:tMatrix}
	\end{align}
	The corresponding equations for $\J = 0$ can easily be obtained by discarding the incident and the emergent wave with $L = \J-1$. The Schr\"odinger equation \eqref{eqn:coupledChannelSE_3x3} is then reduced to two channels and $\mbox{T}_0 = t_{1,1}$.


\section{Including $\bar M_s M_s$ channels}

Since heavy-light and heavy-strange mesons have similar masses, it is essential to also include heavy-strange meson-meson channels $\bar M_s M_s$ in the Schr\"odinger equation \eqref{eqn:Schroedinger_equation} or equivalently \eqref{eqn:coupledChannelSE_3x3}. For example, bottomonium resonances can then be studied for energies as large as $11.025 \, \textrm{GeV}$, the threshold for a negative parity $B$ or $B^\ast$ and a positive parity $B_0^\ast$ or $B_1^\ast$ meson. Without a $\bar M_s M_s$ channel our results would only be valid below the $B_s^{(\ast)} B_s^{(\ast)}$ threshold at $10.807 \, \textrm{GeV}$, which is rather close to the $B^{(\ast)} B^{(\ast)}$ threshold at $10.627 \, \textrm{GeV}$.

	We use the same $2$-flavor lattice QCD static potentials from Ref.\ \cite{Bali:2005fu} to generate the entries of the potential matrix relevant for the $\bar M_s M_s$ channels.	We expect this to be reasonable, since static potentials are known to have a rather mild dependence on light quark masses. This expectation was confirmed by a consistency check with results from a more recent $2+1$-flavor lattice QCD study of string breaking \cite{Bulava:2019iut}. For details we refer to our recent work \cite{Bicudo:2020qhp}.
	
	The coupled channel Schr\"odinger equation projected to definite $\J$ with both $\bar M M$ and $\bar M_s M_s$ scattering channels is given by
	\begin{eqnarray}
		\nonumber & & \hspace{-0.7cm} \left(\frac{1}{2} \mu^{-1} \left(\partial_r^2 - \frac{1}{r^2} L_{\J}^2\right) + V_{\J}(r) \right. \\
		\nonumber & & \hspace{0.675cm} + \left.
		\left(\begin{array}{ccccc}
			E_{\textrm{threshold}} & 0 & 0 & 0 & 0 \\
			0 & 2m_M & 0 & 0 & 0 \\
			0 & 0 & 2m_M & 0 & 0 \\
			0 & 0 & 0 & 2m_{M_s} & 0 \\
			0 & 0 & 0 & 0 & 2m_{M_s} \\
		\end{array}\right) 
		- E\right)
		\left(\begin{array}{c} u_{\J}(r) \\ \chi_{\bar{M}M,\J-1 \rightarrow \J}(r) \\ \chi_{\bar{M}M,\J+1 \rightarrow \J}(r) \\ \chi_{\bar{M_s}M_s,\J-1 \rightarrow \J}(r) \\ \chi_{\bar{M_s}M_s,\J+1 \rightarrow \J}(r) \end{array}\right) = \\
		\nonumber & & = \left(\begin{array}{c} V_{\textrm{mix}}(r) \\ 0 \\ 0 \\ 0 \\ 0 \end{array}\right) \left(\alpha_{\bar{M}M,1} {\J \over 2 \J+1} r j_{\J-1}(k r) + \alpha_{\bar{M}M,2} {\J+1 \over 2 \J+1} r j_{\J+1}(k r)\right. \\
		\label{eqn:coupledChannelSE_5x5} & & \hspace{0.675cm} + \left.\alpha_{\bar{M_s}M_s,1} {1 \over \sqrt{2}} {\J \over 2 \J+1}  r j_{\J-1}(k_s r) + \alpha_{\bar{M_s}M_s,2} {1 \over \sqrt{2}} {\J+1 \over 2 \J+1} r j_{\J+1}(k_s r)\right) ,
	\end{eqnarray}
	with $\mu^{-1} = \textrm{diag}(1/\mu_Q,1/\mu_M,1/\mu_M,1/\mu_{M_s},1/\mu_{M_s})$, $L_\J^2 = \textrm{diag}(\J(\J+1),(\J-1)\J,(\J+1)(\J+2),(\J-1)\J,(\J+1)(\J+2))$ and 
	\begin{align}
		V_\J(r) = 
		\left(\begin{array}{ccccc}
			V_{\bar{Q} Q} & \sqrt{{\J \over 2 \J+1}} V_\textrm{mix}  & \sqrt{{\J+1 \over 2\J+1}} V_\textrm{mix} & {1 \over \sqrt{2}} \sqrt{{\J \over 2 \J+1}} V_\textrm{mix}  & {1 \over \sqrt{2}} \sqrt{{\J+1 \over 2\J+1}}V_\textrm{mix} \\
			\sqrt{{\J \over 2 \J+1}}  V_\textrm{mix}  &0 & 0 & 0 & 0 \\
			\sqrt{{\J+1 \over 2\J+1}}  V_\textrm{mix} & 0  & 0 & 0 & 0 \\
			{1 \over \sqrt{2}} \sqrt{{\J \over 2 \J+1}} V_\textrm{mix}  & 0 & 0 & 0 & 0 \\
			{1 \over \sqrt{2}} \sqrt{{\J+1 \over 2\J+1}} V_\textrm{mix} & 0 & 0 & 0 & 0 \\
		\end{array}\right) .
	\end{align}
	After defining boundary conditions analogously to Eqs.\ \eqref{eqn:boundary_conditions_1} to \eqref{eqn:boundary_conditions_3}, we obtain a 4x4-scattering matrix
	\begin{eqnarray}
		\nonumber & & \hspace{-0.7cm} \mbox{T}_\J =
		\left(\begin{array}{cccc}
			t_{\bar{M}M, \J-1; \bar{M}M, \J-1} & t_{\bar{M}M, \J+1; \bar{M}M, \J-1}  & t_{\bar{M_s}M_s, \J-1; \bar{M}M, \J-1} & t_{\bar{M_s}M_s, \J+1; \bar{M}M, \J-1} \\
			t_{\bar{M}M, \J-1; \bar{M}M, \J+1} & t_{\bar{M}M, \J+1; \bar{M}M, \J+1}  & t_{\bar{M_s}M_s, \J-1; \bar{M}M, \J+1} & t_{\bar{M_s}M_s, \J+1; \bar{M}M, \J+1} \\
			t_{\bar{M}M, \J-1; \bar{M_s}M_s, \J-1} & t_{\bar{M}M, \J+1; \bar{M_s}M_s, \J-1}  & t_{\bar{M_s}M_s, \J-1; \bar{M_s}M_s, \J-1} & t_{\bar{M_s}M_s, \J+1; \bar{M_s}M_s, \J-1} \\
			t_{\bar{M}M, \J-1; \bar{M_s}M_s, \J+1} & t_{\bar{M}M, \J+1; \bar{M_s}M_s, \J+1}  & t_{\bar{M_s}M_s, \J-1; \bar{M_s}M_s, \J+1} & t_{\bar{M_s}M_s, \J+1; \bar{M_s}M_s, \J+1} \\
		\end{array}\right) . \\
		\label{T_5x5} & &
	\end{eqnarray}


\section{Results}

	Now we focus on heavy $b$ quarks, take the $b$ quark mass from quark models ($m_Q = 4.977 \, \textrm{GeV}$ \cite{Godfrey:1985xj}) and use the spin averaged mass of the $B$ and the $B^\ast$ meson ($m_M = (m_B + 3 m_{B^\ast}) / 4 = 5.313 \, \textrm{GeV}$) and of the $B_s$ and the $B_s^\ast$ meson ($m_{M_s} = (m_{B_s} + 3 m_{B_s^\ast}) / 4 = 5.403 \, \textrm{GeV}$). The lattice QCD data we are using \cite{Bali:2005fu} corresponds to $E_{\textrm{threshold}} = 10.790 \, \textrm{GeV}$.

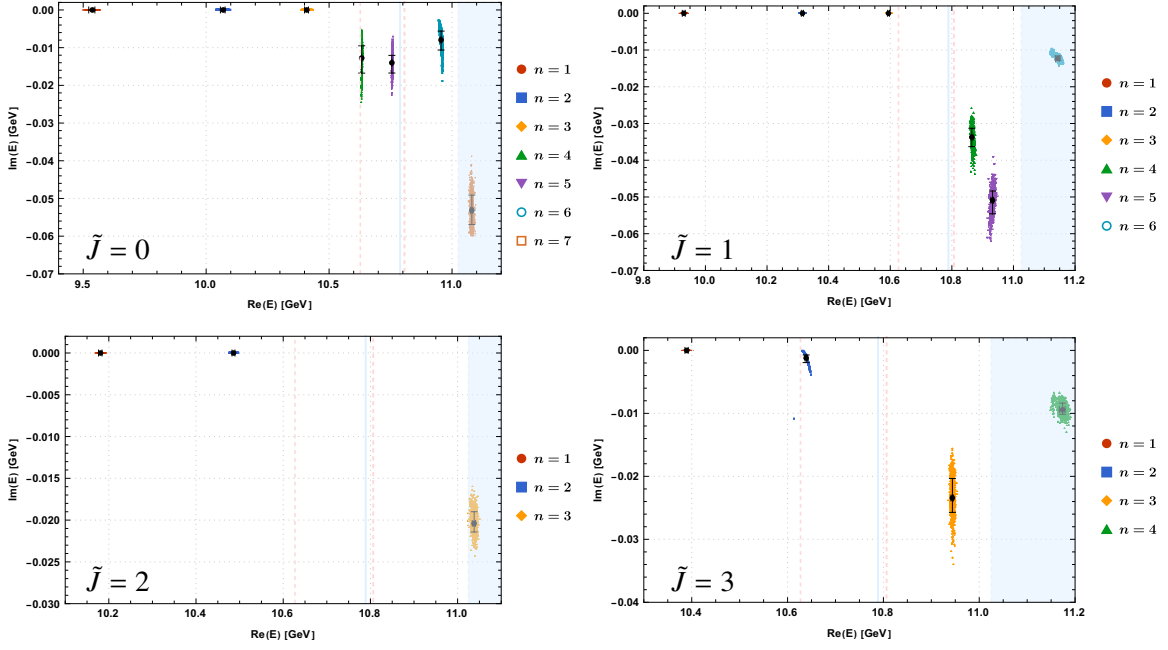
\begin{figure}
		\centering
		\tikzmath{\x = 3.85; \y = 2.2; \imagescale = 0.5; \titlex = 2.3; \titley = 1.2;}
		\begin{tikzpicture}
			\node[inner sep=0] (image1) at (-\x,\y) {\includegraphics[width=\imagescale\textwidth]{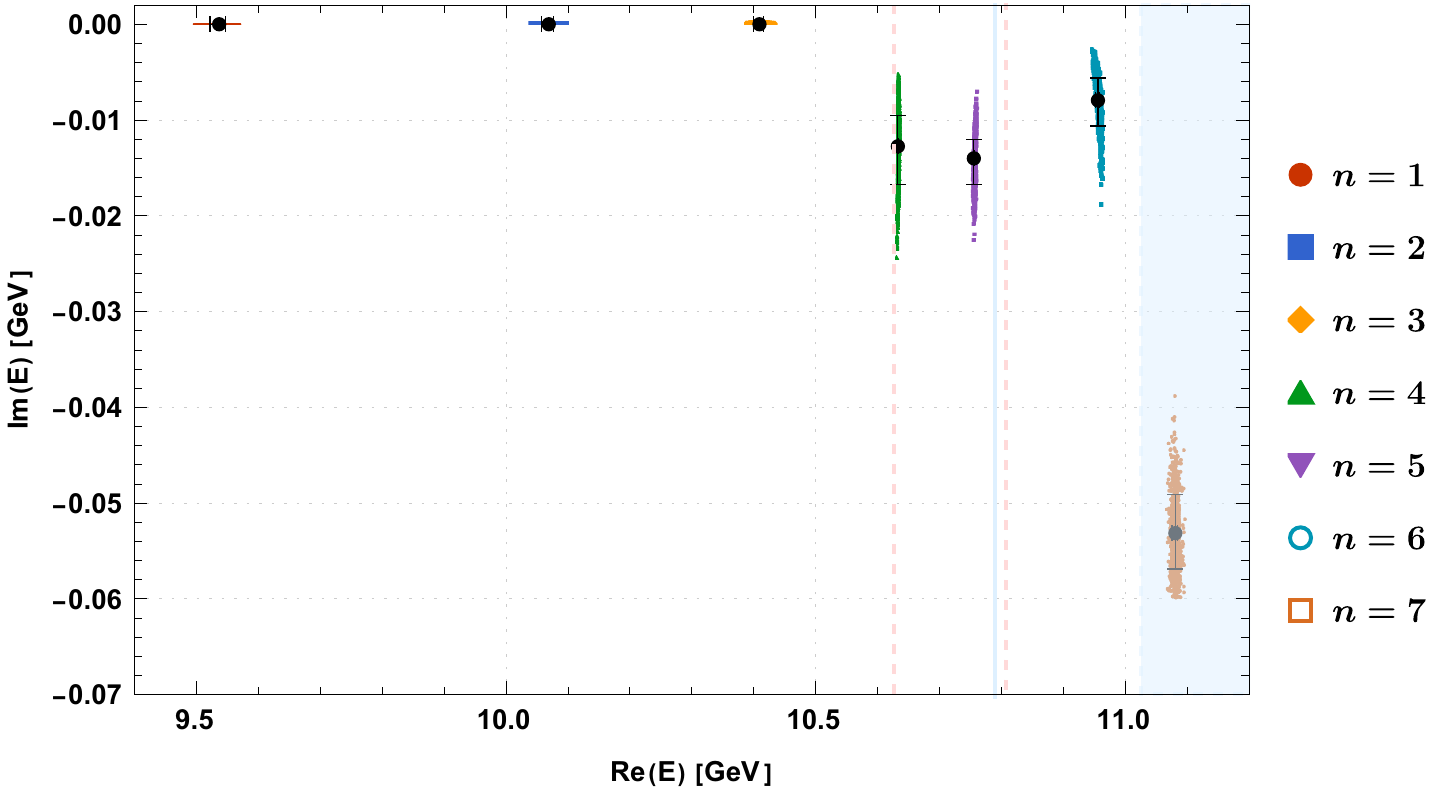}};
			\node at ($(image1)-(\titlex,\titley)$) {$\J = 0$};
			\node[inner sep=0] (image2) at (\x,\y) {\includegraphics[width=\imagescale\textwidth]{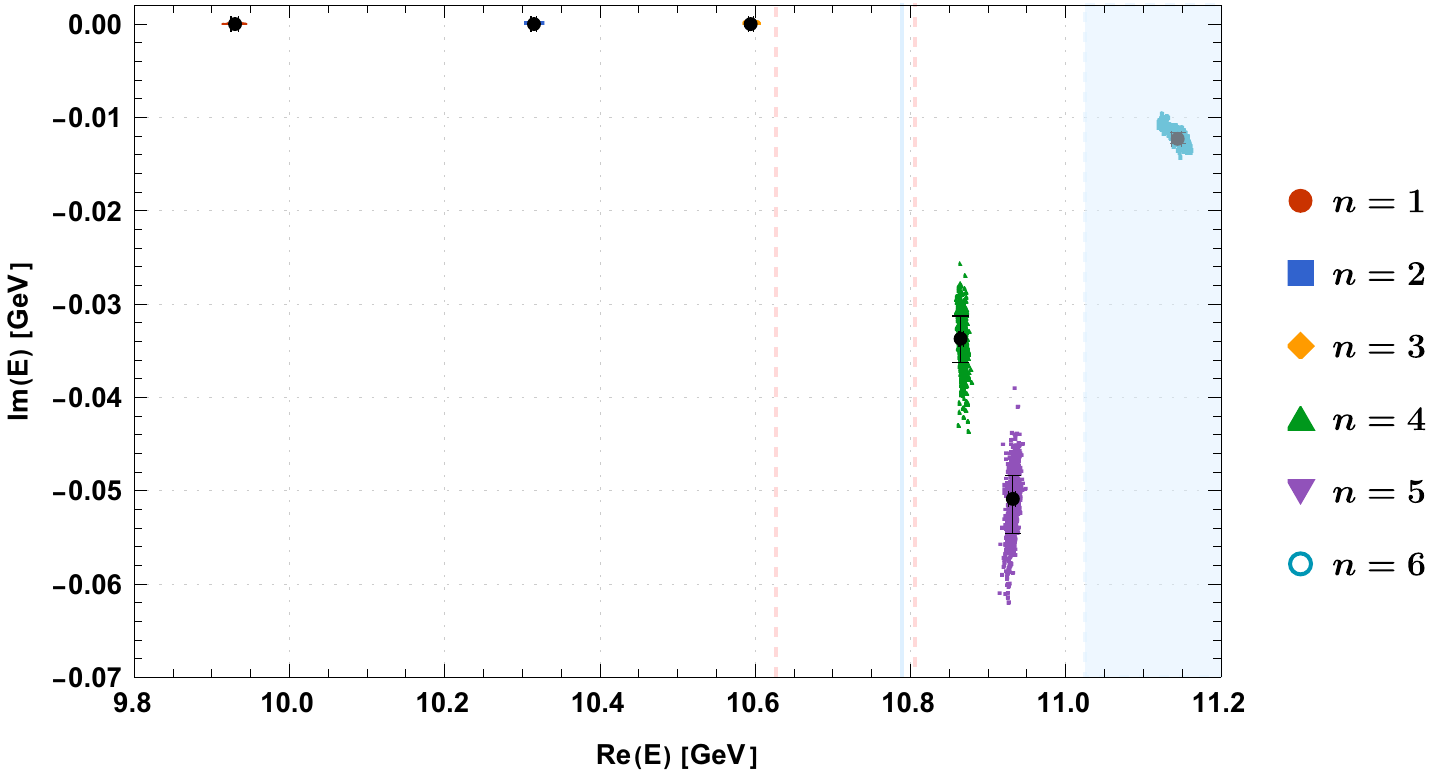}};
			\node at ($(image2)-(\titlex,\titley)$) {$\J = 1$};
			\node[inner sep=0] (image3) at (-\x,-\y) {\includegraphics[width=\imagescale\textwidth]{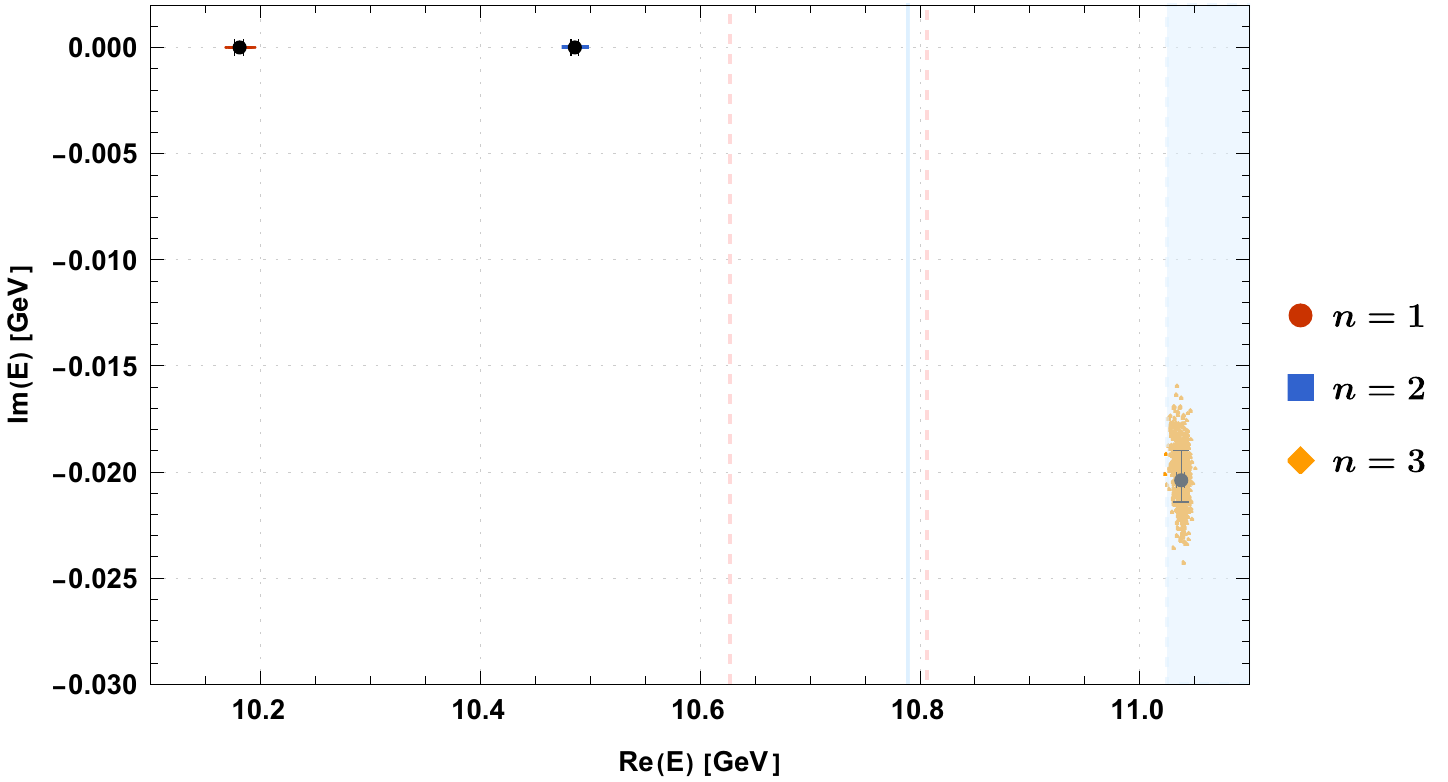}};
			\node at ($(image3)-(\titlex,\titley)$) {$\J = 2$};
			\node[inner sep=0] (image4) at (\x,-\y) {\includegraphics[width=\imagescale\textwidth]{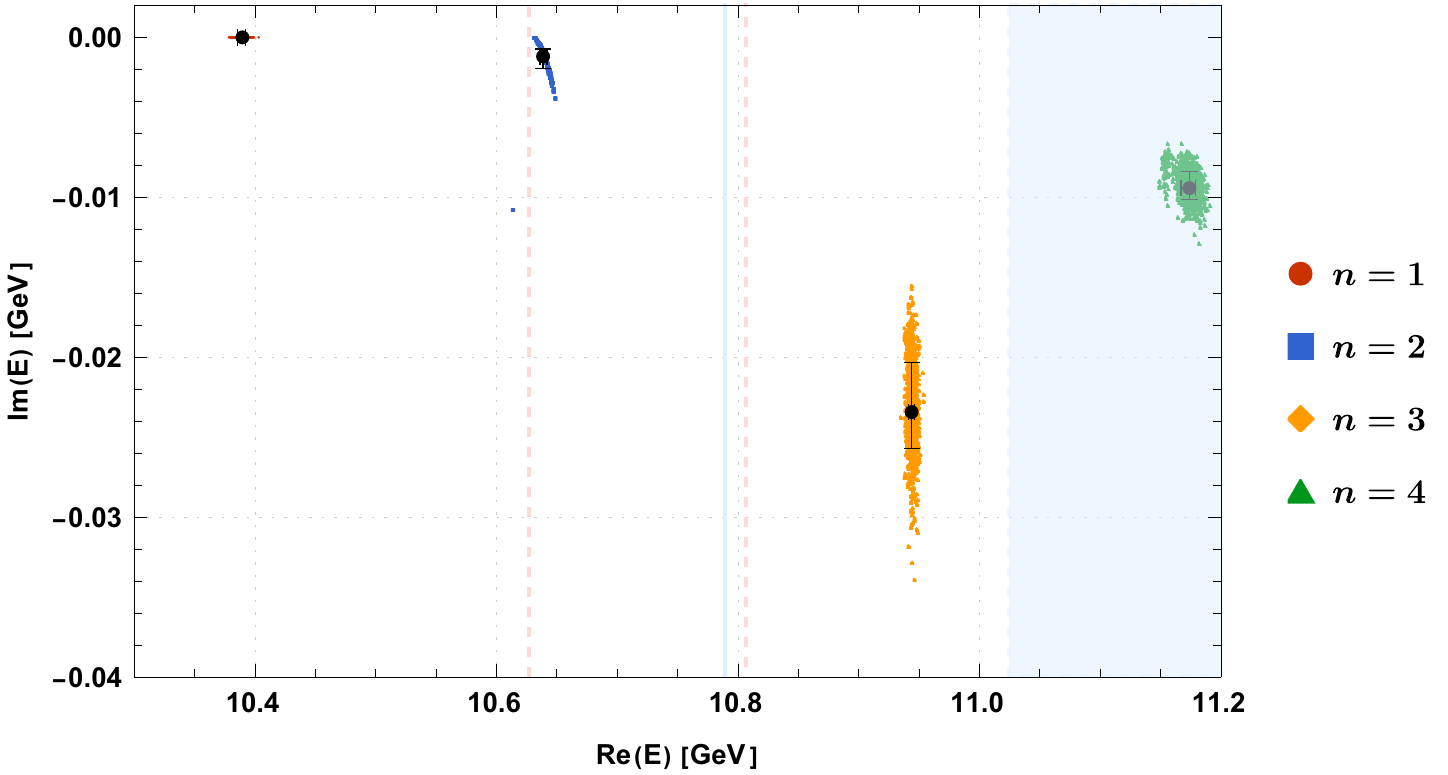}};
			\node at ($(image4)-(\titlex,\titley)$) {$\J = 3$};
		\end{tikzpicture}
		\caption{Poles of $\mbox{T}_\J$ in the complex plane representing bound states and resonances below $11.2 \, \textrm{GeV}$. Colored point clouds and the corresponding black error bars reflect statistical errors of the lattice QCD data from Ref.\ \cite{Bali:2005fu}. The vertical dashed lines indicate the $\bar B^{(*)} B^{(*)}$ and $\bar B^{(*)}_s B^{(*)}_s$ thresholds. The light blue shaded regions above $11.025 \, \textrm{GeV}$ mark the opening of the threshold of one heavy-light meson with negative parity and another one with positive parity. Results in these regions should not be trusted anymore.}
		\label{fig:polepositions}
	\end{figure}

	We consider the analytic continuation of the coupled channel Schr\"odinger equation \eqref{eqn:coupledChannelSE_5x5} to the complex energy plane, where we search for poles of the $\mbox{T}$ matrices \eqref{T_5x5}. In Figure~\ref*{fig:polepositions} we show all poles with real part below $11.2 \, \textrm{GeV}$. To propagate the statistical errors of the lattice QCD data, we generated 1000 statistically independent samples and repeated our computations on each of these samples. For each bound state and resonance there is a differently colored point cloud representing the 1000 samples. Bound states are located on the real axis below the $\bar B^{(*)} B^{(*)}$ threshold at $10.627 \, \textrm{GeV}$, while resonances are above this threshold and have a non-vanishing negative imaginary part. As usual, the pole energies $E$ are related to masses and decay widths of bottomonium states via $m = \textrm{Re}(E)$ and $\Gamma = -2\,\textrm{Im}(E)$.
	
	In Table \ref{tab:results} we compare our results with experimentally observed bound states and resonances.

	The low-lying states we found have masses similar to those measured in experiments and it seems straightforward to identify their counterparts:
	\begin{itemize}
	\item $\J = 0$, $n=1,2,3,4$ correspond to $\eta_b(1S) \equiv \Upsilon(1S)$, $\Upsilon(2S)$, $\Upsilon(3S)$ and $\Upsilon(4S)$.
	\item $\J = 1$, $n=1,2,3$ correspond to $h_b(1P) \equiv \chi_{b0}(1P) \equiv \chi_{b1}(1P) \equiv \chi_{b2}(1P)$, $h_b(2P) \equiv \chi_{b0}(2P) \equiv \chi_{b1}(2P) \equiv \chi_{b2}(2P)$ and $\chi_{b1}(3P)$.
	\item $\J = 2$, $n=1$ corresponds to $\Upsilon(1D)$.	
	\end{itemize}

	Our resonance mass for $\J = 0$, $n = 5$ is quite similar to the experimental result for $\Upsilon(10753)$, which was recently reported by Belle \cite{Belle:2019cbt}. In a recent publication \cite{Bicudo:2020qhp} we investigated the structure of this state within the same setup and found that it is meson-meson dominated. Thus, since it is not an ordinary quarkonium state and the heavy quark spin can be $1^{--}$, it can be classified as a $Y$ type crypto-exotic state.

	The resonances $\Upsilon(10860)$ and $\Upsilon(11020)$ are typically interpreted as $\Upsilon(5S)$ and $\Upsilon(6S)$. However, from the experimental perspective they could as well correspond to $D$ wave states. Since our $\J = 0$, $n = 6$ state is rather close to $\Upsilon(10860)$, while there is no matching candidate for $\J = 2$, our results clearly support the interpretation of $\Upsilon(10860)$ as $\Upsilon(5S)$. Concerning $\Upsilon(11020)$ the situation is less clear. First, the mass of $\Upsilon(11020)$ is already close to the threshold for a negative parity $B$ or $B^\ast$ and a positive parity $B_0^\ast$ or $B_1^\ast$ meson, a channel we have not yet included in our approach. Second, the $\J = 0$, $n = 7$ and $\J = 2$, $n = 3$ states have almost the same mass and are both close to $\Upsilon(11020)$. Thus, we are not in a position to decide, whether the $\Upsilon(11020)$ is an $S$ wave or rather a $D$ wave state.

	\begin{table}
		\centering
		\includegraphics[width=0.7\textwidth]{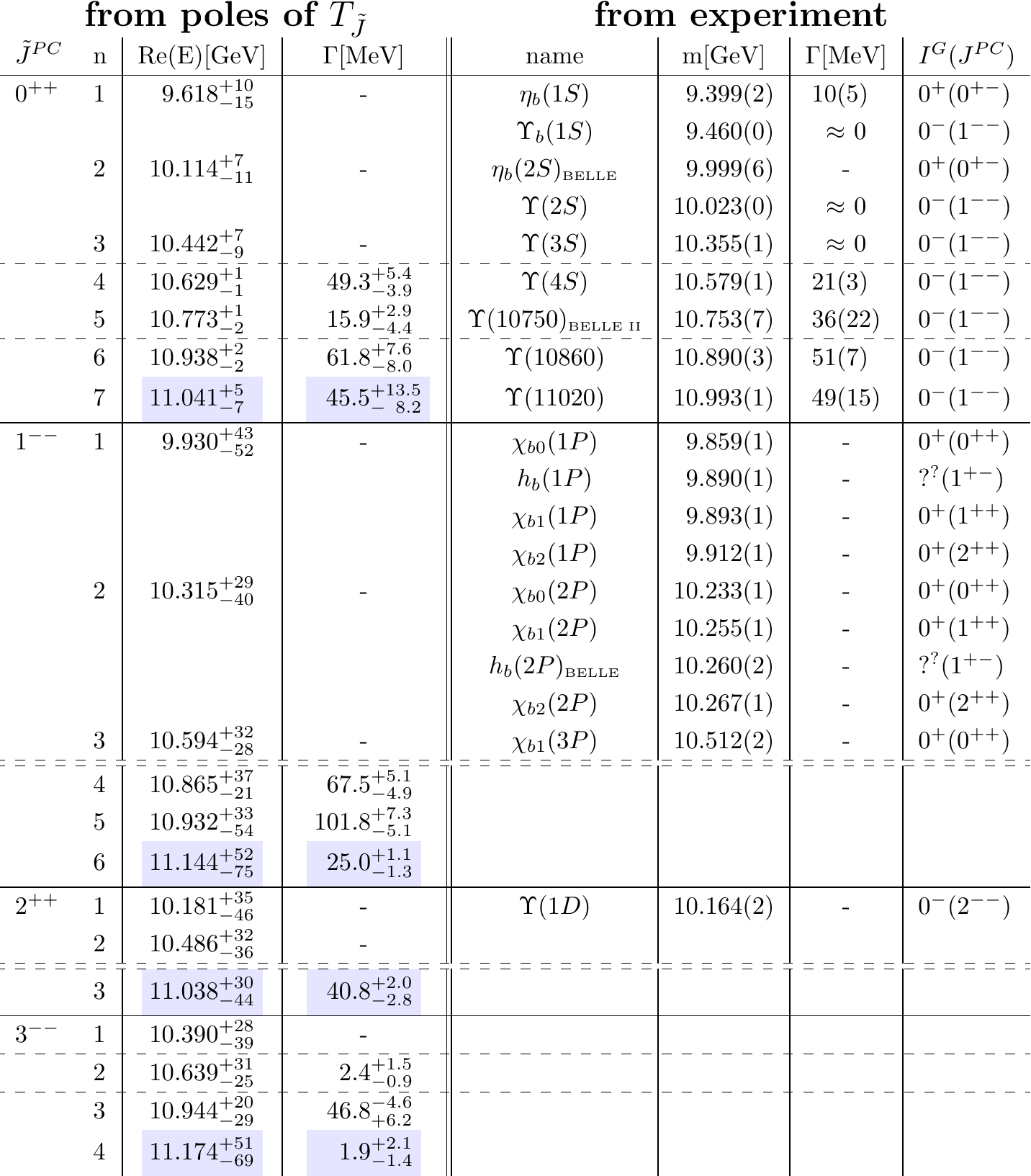}
		\caption{Masses and decay widths for $I = 0$ bottomonium from the coupled channel Schr\"odinger equation \eqref{eqn:coupledChannelSE_5x5}. For comparison we also list available experimental results. The $\bar B^{(*)} B^{(*)}$- and $\bar B^{(*)}_s B^{(*)}_s$-threshold are marked by dashed lines. Errors on our results are purely statistical.}
		\label{tab:results}
	\end{table}

	Since we neglected effects due to the heavy quark spins, we expect that our results on have systematic errors of order $50 \, \textrm{MeV}$. Including heavy quark spins in our approach is a major step \cite{Bicudo:2016ooe}, which we plan to take in the near future.
	

\section*{Acknowledgements}

We acknowledge useful discussions with Gunnar Bali, Eric Braaten, Marco Cardoso, Francesco Knechtli, Vanessa Koch, Sasa Prelovsek, George Rupp and Adam Szczepaniak. L.M. acknowledges support by a Karin and Carlo Giersch Scholarship of the Giersch foundation. M.W.\ acknowledges funding by the Heisenberg Programme of the Deutsche Forschungsgemeinschaft (DFG, German Research Foundation) -- Projektnummer 399217702. Calculations on the Goethe-HLR and on the FUCHS-CSC high-performance computer of the Frankfurt University were conducted for this research. We would like to thank HPC-Hessen, funded by the State Ministry of Higher Education, Research and the Arts, for programming advice. PB and NC thank the support of CeFEMA under the contract for R\&D Units, strategic project No. UID/CTM/04540/2019, and the
FCT project Grant No. CERN/FIS-COM/0029/2017. NC is supported by FCT under the contract No. SFRH/BPD/109443/2015.


\end{document}